\documentclass[preprint, amssymb, amsmath, nobibnotes, nofootinbib, aps, prb,showpacs]{revtex4-1}
\usepackage[dvips]{graphicx}
\begin{document}
\title{Covalency driven modulation of paramagnetism and development of lone pair ferroelectricity in multiferroic Pb$_3$TeMn$_3$P$_2$O$_{14}$}
\author{Rafikul Ali Saha$^1$}
\author{Anita Halder$^2$}
\author{Tanusri Saha-Dasgupta$^{2}$}
\author{Desheng Fu$^3$}
\author{Mitsuru Itoh$^4$}
\author{Sugata Ray$^{1}$}
\email{mssr@iacs.res.in}
\affiliation{$^1$School of Materials Sciences, Indian Association for the Cultivation of Science, 2A \& 2B Raja S. C. Mullick Road, Jadavpur, Kolkata 700 032, India}
\affiliation{$^2$Department of Condensed Matter Physics and Material Sciences, S. N. Bose National Centre for Basic Sciences, Block JD, Sector 3, Saltlake, Kolkata -700106, India}
\affiliation{$^3$Department of Electronics and Materials Science, and Department of Optoelectronics and Nanostructure Science, Graduate School of Science and Technology, Shizuoka University, 3-5-1 Johoku, Naka-ku, Hamamatsu 432-8561, Japan}
\affiliation{$^4$Materials and Structures Laboratory, Tokyo Institute of Technology, 4259 Nagatsuta, Yokohama 226-8503, Japan}
\pacs {}
\begin{abstract}
We have investigated the structural, magnetic and dielectric properties of Pb-based langasite compound Pb$_3$TeMn$_3$P$_2$O$_{14}$ both experimentally and theoretically in the light of metal-oxygen covalency, and the consequent generation of multiferroicity. It is known that large covalency between Pb 6$p$ and O 2$p$ plays instrumental role behind stereochemical lone pair activity of Pb. The same happens here but a subtle structural phase transition above room temperature changes the degree of such lone pair activity and the system becomes ferroelectric below 310 K. Interestingly, this structural change also modulates the charge densities on different constituent atoms and consequently the overall magnetic response of the system while maintaining global paramagnetism behavior of the compound intact. This single origin of modulation in polarity and paramagnetism inherently connects both the functionalities and the system exhibits mutiferroicity at room temperature.
\end{abstract}
\maketitle

${Introduction}$. The story of stereochemically active cationic lone pairs, arising due to $s$-$p$ mixing in metal assisted by covalency with the ligand $p$-orbitals and finally driving ferroelectricity within a polar unit cell, has always been an exciting point of discussion in condensed matter physics~\cite{Walsh}. There is a sizable number of systems where this mechanism plausibly works, such as, ferroelectric perovskites (PbTiO$_3$, BiMnO$_3$, BiFeO$_3$, SnTiO$_3$, CsPbF$_3$ etc)~\cite{Jin, Hill, Seshadri, Volkova, Nakhmanson, Fennie}, double perovskites (Pb$_2$ScTi$_{0.5}$Te$_{0.5}$O$_6$, Pb$_2$ScSc$_{0.33}$Te$_{0.66}$O$_6$, Pb$_2$MnWO$_6$)~\cite{Alonso, Mathieu}, as well as $\alpha$-PbO~\cite{Watson}, SnO, BiOF~\cite{Ramseshadri}, Bi$_2$WO$_6$~\cite{Stolen}, BiMn$_2$O$_5$~\cite{Shukla} etc. Among these, systems which also accommodate magnetic cations and consequent possibility of multiferroicity draws much enhanced attention. The two most prominent examples of such multiferroics having significant roles of lone pair and covalency are BiFeO$_3$~\cite{Volkova, Park} and BiMnO$_3$~\cite{Seshadri, Seshadri2}. Among these, BiFeO$_3$ undergoes ferroelectric phase transition with $T_C$~=~1100 K~\cite{Smith} and a G-type antiferromagnetic transition at a lower temperature (650 K), accompanied by an incommensurate spin cycloid structure having a period of 620 \AA~\cite{Sosnowska, Przenioslo}. Due to this incommensurate spatially modulated cycloid spin structure, linear ME effect is not observed in BiFeO$_3$, however, the polarization breaks the crystal symmetry and manifests itself in the appearance of an inhomogeneous ME interaction (Lifshitz invariant), showing quadratic dependence of polarization on the magnetic field~\cite{Kadomtseva}. On the other hand, BiMnO$_3$ is one of the very few multiferroics that is ferroelectric ($\sim$~770~K) and ferromagnetic ($T_C$~=~105~K)~\cite{Luca, Chou, Chi}. In this case the ferroelectricity has been explained by the presence of stereochemically active Bi$^{3+}$ 6$s^2$ lone pair, developing from the mixing of the Bi 6$s$ and 6$p$ orbitals, activated by charge transfer from fully filled anionic orbital to the empty Bi 6$p$ orbital. As a result, the system becomes structurally noncentrosymmetric below 770~K~\cite{Kimura} and a distinct magnetoelectric response of -0.6~\% is observed at the magnetic transition temperature~\cite{Kimura}.
\par
In this context we have explored rarely studied langasites ($A_3$$B$$C_3$$D_2$O$_{14}$), having lone pair bearing Pb$^{2+}$ ions and 3$d$ transition metal (TM) ions within TMO$_4$ tetrahedral units. Tetrahedral co-ordinations are traditionally more conducive for TM-O covalent interaction~\cite{Armstrong, Helmholtz, Lee, Woodword} which plays a crucial role in deciding the degree of lone pair activity of Pb by manipulating the available electron density on the oxygen, sharing both TM and Pb connections. In the family of langasites, $C$ and $D$ site cations remain in tetrahedral coordination, while $B$ and $A$ sites form octahedral and decahedral network respectively~\cite{Marty, Krizan, Silverstein1, Silverstein2, Markina1, Markina2}. The network of $A$ cations forms topologically equivalent kagome lattice from well separated planes of corner sharing triangles~\cite{Marty2}. Members of Langasite family are known to host multiferroism, non-linear optical, pizoelectric, ferroelectric and dielectric properties~\cite{Krizan}, e.g., Fe based langasite compounds $A_3$$B$Fe$_3$$D_2$O$_{14}$ ($A$ = Ba, Sr, Ca; $B$ = Ta, Nb, Sb; $D$ = Ge, Si) show frustration driven helical magnetic order, accompanied by a dielectric anomaly at the onset of magnetic transition~\cite{Marty, Marty1}. However, lone pair driven multiferroicity in langasites has not been explored till date.
\par
Pb$_3$TeMn$_3$P$_2$O$_{14}$ can become important exactly in this context where changes in the degree of stereochemical lone pair activity of Pb$^{2+}$ distorts the general non-polar structure into a polar one, causing a shifting of magnetic Mn/P planes (M-P planes) towards the neighbouring Pb/Te containing planes (P-T plane) in the direction of positive $c$ axis. This results in formation of stripe-like distribution of pairwise closely placed M-P / P-T planes within the structure which in turn enhances the covalency as well as magnetic moments on otherwise nonmagnetic entities such as Pb$^{2+}$, Te$^{6+}$, P$^{5+}$, and O$^{2-}$. This structural change brings forth spontaneous polarization in the system at around $\sim$310~K, and the synchronous redistribution of magnetic moments on the ions and shortened relative distances among them affect the overall paramagnetic state, which gets manifested in a clear signal of magnetoelectric coupling within the apparent paramagnetic phase, {\it i.e.} far above the long range magnetic transition ($\sim$~7~K) of the system. Such unusual phenomena has been reported recently~\cite{Sampathkumaran1, Sampathkumaran2} in spin-chain compounds and has been explained by the variations of short-range magnetic correlations within the paramagnetic phase.
\par
Overall, in this paper we present a new multiferroic with room temperature magnetoelectric coupling arising from covalency affected stereochemical activity of lone pair and coincident changes in local magnetic correlations.

Experimental and theoretical details are elaborately presented in the Supplemental Material~\cite{Supplementary}.

$Structure$ $from$ $x-ray$ $diffraction$. The structural refinement of room temperature X-Ray diffraction (XRD) of PTMPO has been performed by considering noncentrosymmetric and polar trigonal space group ($P$3) (Lattice parameter and crystal structure are given in Table S1 and S2 of the Supplemental Material~\cite{Supplementary}), consistent with previous literature reports~\cite{Silverstein3, Krizan}. Temperature dependent X-ray diffractions have been collected over a wide temperature range of 5-400 K. The Rietveld refinements~\cite{Supplementary} of these collected XRD patterns have been carried out using the same trigonal space group $P$3. Thermal variations of refined lattice parameters, $a$, $b$, $c$ and unit cell volume, are shown in Fig. 1 (a)-(c). Two anomalies at $\sim$~310~K and $\sim$~120~K are observed in the temperature dependent lattice parameters as well as unit cell volume variations. We performed the 299 K XRD refinement using both $P$321 and $P$3 space group, as shown in Fig. S1 (a) and (b) of the Supplemental Material~\cite{Supplementary} respectrively. It is clearly evident from the fitting that the higher symmetry $P$321 space group could not capture the supper lattice peaks in the low temperature diffraction pattern (blue ellipse line in Fig. S1 (a) of Supplemental Material~\cite{Supplementary}). On the other hand, the super lattice peaks have been well reproduced while fitting using low symmetry $P$3 space group (as indicated in Fig. S1 (b) of the Supplemental Material~\cite{Supplementary}). Most importantly, the room temperature synchrotron data also are very similar to our lab XRD results, as shown in Fig. S1 (d), (e) and (f) of the Supplemental Material~\cite{Supplementary}. This convincingly affirms the room temperature ferroelectric phase ($P$3) for the studied sample. It is important to note that the absence of superlatticepeaks in the 400 K XRD refinement (see Fig. S1 (c) of the Supplemental Material~\cite{Supplementary}) signifies the higher symmetric structure (space group: $P$321) of the sample above 310 K which is noncentrosymmetric and non polar. Therefore, it can be concluded that above 310 K, the system has a non-polar $P$321 structure (see Table S1 and S3 of the Supplemental Material~\cite{Supplementary}) but the symmetry gets lifted (polar $P$3) with cooling and the polar structure only gets further distorted below 120~K.

Due to this structural phase transition ($P$321 to $P$3), the unit cell becomes larger and all Pb and Mn equilateral triangles get converted to scalene triangles (see Fig. S1 (g) of the Supplemental Material~\cite{Supplementary}). The lower symmetry $P$3 space group is a sub group of $P$321, and is achieved by removing the $C_2$ symmetry from $P$321. This happens when the M-P plane is shifted towards the P-T plane in the direction of positive $c$ axis (Fig. 2(a)-(b)). As a result, Mn-O bond lengths become shorter ensuring larger covalent interaction (see the Table I) and plausible enhancement of the stereochemical activity of the Pb$^{2+}$ lone pair. This shift also affects the overall site symmetry as all the Pb and Mn triangles become scalene, as shown in Fig. 2(c)-(d). This unit cell with lower symmetry becomes large, containing seven formula units per unit-cell.
Interestingly, these scalene triangles of Pb, Mn, and Te are found to be connected through a systematic 120$^\circ$ rotations (see Fig. S2 and Fig. S3 of the Supplemental Material~\cite{Supplementary}) and it may be anticipated that stereochemically active Pb$^{2+}$ (6$s^2$) lone pair is responsible for such deformations of the structural motifs.

However, it is primarily important to confirm the existence of stereochemically active Pb$^{2+}$ lone pair at all temperatures, while the degree may change with structural transitions. If we divide the Pb-O polyhedron into two spheres (I and II), the difference between the shortest Pb-O distances of the two spheres ($\triangle$$E_1$) provides a measure of the asymmetry, degree of delocalization and lone pair activity. The other notable parameter in this regard is $\triangle$$E_2$ which is the difference between the shortest Pb-O distance in the polyhedron of PTMPO and the known shortest Pb-O distance in compounds containing these elements. A large value of $\triangle$$E_1$ and smaller value of $\triangle$$E_2$ attribute higher stereochemical activity of the lone pair~\cite{Volkova}. Here, to estimate $\triangle$$E_2$, we considered the minimal distance $d_{Pb^{2+}-O}$ (2.40 \AA)~from Pb$_2$V$_5$O$_{12}$~\cite{Volkova}. Thus the values of $\triangle$$E_1$ turn out to be 0.44 \AA~and 0.36 \AA~while $\triangle$$E_2$ to be -0.04 \AA~and 0.06 \AA~for $P$3 and $P$321 space groups respectively, indicating higher degree of stereochemical activity of Pb$^{2+}$ lone pair in the $P$3 structure of PTMPO, which is in line with the existing results of BiFeO$_3$ compounds~\cite{Volkova}.

$Theoretical$ $calculations$. We have carried out density functional theory (DFT)~\cite{Supplementary} calculations in order to assess the covalency effect which is important for the stereochemical activity of the Pb lone pair. Since the goal of our first-principles calculation is to reveal the covalency effect between different cations and oxygen and the effect of structural transition on it, the analysis of electron structure was carried out on spin-polarized calculations. The antiferromagnetic structure, which is the ground state magnetic structure is complex and leads to cancellation of covalency effect due to opposite alignment of spins, as ascertained in vanishing moment of oxygen compared to its finite value in fully spin-polarized calculation (see Table II). The calculations have been carried out both for the nonpolar (high temperature $P$321) and polar (low temperature $P$3) structures and the detailed density of states (DOS) are shown in Fig. 3(a) and (d), respectively. The stereochemical activity of lone pair active ions like Bi$^{3+}$ or Pb$^{2+}$ within oxide structures has been argued to originate from the hybridization between 6$s$, 6$p$ orbitals at Bi/Pb site and 2$p$ orbitals at O site. The degree of stereochemical activity is thus crucially dependent on the energy level positions of these states, especially the position of  6$p$. The 6$p$-2$p$ hybridization turns out to be governing factor in giving rise to directionality of the lone pair and driving the off-centric movement~\cite{Walsh}. Comparing between the projected DOS of PTMPO at high temperature non polar ($P$321) and low temperature polar  ($P$3) phases we observe a marked change in the position of Pb 6$p$, it being closer to O 2$p$ by about 1.5 eV in the low temperature phase, compared to that in the high temperature phase, suggestive of enhanced interaction between Pb 6$s$ - O 2$p$ antibonding orbitals with the empty Pb 6$p$ orbital~\cite{Walsh}. This has been quantitatively confirmed from our COHP (Crystal Orbital Hamilton Population)~\cite{Supplementary} calculation where the integrated COHP (ICOHP) values have been found to become double for the Pb-O connectivity while going from $P$321 to $P$3 structure (see Fig. S4 and Table S4 in the Supplemental Material~\cite{Supplementary}). Such an effect should obviously enhance the degree of Pb lone pair activity in the compound at lower temperature. Interestingly, in the langasite structures, this also affects the energy position of the orbitals and electron density on Te because Pb and Te share common oxygen. This effect is clearly reflected in projected DOS, which shows Te 5$p$ to be closer to oxygen 2$p$ by 1 eV in the low temperature phase compared to that in high temperature phase. As a result of this, there evolves an off-centric movement of otherwise nominally Te$^{6+}$ ion with two well defined coordination shells of neighboring oxygen atoms in the low temperature phase (see Fig. 3. (f)), while at high temperature the off centric movement of Te vanishes with only one coordination shell of neighboring oxygen atoms (see Fig. 3. (c)). Thus, clearly dipoles are created within the system.


On top of it, the elemental magnetic moment possesses a noticeable change in the polar structure relative to the high temperature nonpolar structure, in line with the stronger covalency effect of the polar structure (see table II). The charge density distribution in the two structures indicate that there could be one more dipole moment center that might be created in the polar structure ($P$3) other than the polar geometry of TeO$_6$ octahedra. The  stereochemically active lone pairs of Pb$^{2+}$ ($ns^2$) in Pb-Pb hexagon do not show any net polarization due to the equal length of Pb-Pb distance in the $P$321 phase, as shown in Fig. 3. (c), while in the polar structure, Pb$_6$ hexagon creates a net dipole moment along the negative $b$ direction with respect to the centre of mass, as shown in Fig. 3. (f). As Pb$_6$ hexagon and TeO$_6$  octahedra posses dipole moments along the negative $b$ and negative $c$ directions, respectively, a net dipole moment within a single Pb$_6$ motif (where Te and Pb are present in the a-b plane) is generated along the b-c plane.

$Dielectric$ $measurements$. All the results shown above indicate towards realizing a ferroelectric phase at room temperature in PTMPO which is next checked by dielectric measurements. The temperature dependence of the real part of the dielectric constant ($\varepsilon'$/$\varepsilon_0$) and dielectric loss ($\tan\delta$)  at different frequencies (1 kHz to 1 MHz) in the temperature range of 5 to 400 K are shown in Fig. 4(a) and (b). A glass like phase transition (frequency dependent) near 120 K and another frequency independent anomaly  near 315 K appears in the data.  Clearly, the anomaly near 315 K appears due to the structural phase transition from high temperature nonpolar phase $P$321 to low temperature polar phase $P$3. Further the polarization ($P$) with varying electric field ($E$) at room temperature (298 K) has been measured, as indicated in Fig. 4(c), where
clear $P-E$ loop with saturation and remnant polarization of 0.83 $\mu$C/$cm^2$ and 0.3 $\mu$C/$cm^2$ respectively, are observed at room temperature. Such a square loop along with sharp peaks in the switching current density ($J$) vs electric field ($E$) curve at room temperature (Fig. 4(c)(sky shaded region)), clearly confirm existence of ferroelectric ordering in the sample at room temperature.

$Magnetism,$ $heat$ $capacity$ $and$ $magnetodielectric$. Next we focus on the magnetic properties of the compound. Zero-field cooled (ZFC), field cooled cooling (FCC) and field cool heating (FCH) magnetization at 500 Oe, 5000 Oe and 10000 Oe were measured in the temperature range of 2-400 K, as shown in Fig. 5(a). Clear antiferromagnetic transition around $T_N$ = 7 K is observed in all the ZFC, FCC and FCH curves, consistent with previous studies~\cite{Silverstein1, Krizan}.

Interestingly, an unexpected effect is seen in the magnetic susceptibility at high temperature around the nonpolar to polar structural phase transition. A wide and thin hysteresis between the FCC and FCH data is observed around room temperature along with a distinct feature in the FCH curve (Fig. 5(b)). Such a behavior is reminiscent of a first order phase transition which is unusual for a purely paramagnetic state. In order to reconfirm, we have repeatedly measured the $M-T$ data of different batches of samples in different instruments and obtained the same result confirming the robustness of the observation. It has been discussed earlier that the shift of the M-P plane towards P-T plane at the phase transition, affects the local covalency, effective moments on all the atoms (see Table-I and II), and probably the local magnetic correlations as well. It is likely that the local magnetic interaction gets affected quite substantially even though the system continues to remain a paramagnet globally. We tend to express the two different paramagnetic phases as PM-II (polar phase) and PM-I (nonpolar phase) which must be differing at the local scale as indicated by the susceptibility data.
These features in magnetic susceptibility are further characterized by the zero field heat capacity measurements ($C_p$ versus $T$). A sharp $\lambda$ like anomaly, authentication mark of thermodynamic phase transition into a long range magnetic ordering has been observed near 7 K in the $C_p$ versus $T$ data (shown in the inset of Fig. 5 a)) which is in agreement with magnetic susceptibility data. Interestingly, we observe an anomaly just above room temperature (see the inset of Fig. 5 b)), which maps the high temperature magnetic anomaly (bifurcation in FCC and FCH) and the structural phase transition.

Further we study the temperature dependent dielectric constant under 9 Tesla magnetic field, shown in Fig. 5(c) along with the zero field data. A significant change is observed between the with field ($H$ = 9T) and without field data near 270 K, as indicated in the inset to Fig. 5(c), signifying the presence of magnetoelectric coupling in the system. This magnetoelectric coupling is further established from the isothermal magnetoelectric data at 270 K, as shown in Fig. 5(d). Further, there is no significant change in magnetocapacitance with varying magnetic field at 315 K (see Fig. S5 of the Supplemental Material~\cite{Supplementary}) signifying the absence of magnetoelectric coupling above ferroelectric phase transition. It can be argued that the structural phase transition brings forth ferroelectricity in the system below 310 K as a result of enhanced covalency and lone pair activity which at the same time affects the magnetism locally (introduction of the PM-II phase) and ensures a definite coupling between the two as they originate from the same microscopic effect.

$Conclusion$. We have reported the results of experimental and theoretical studies on Pb based langasite compound PTMPO. A clear ferroelectric transition is developed near 310 K. Stronger Mn-O covalent interactions helps to redistribute the charges among the other cation-oxygen bonds, which in turn induces crucial electronic changes and consequently, the individual elemental magnetic moment gets changed in the system. The Mn-O covalency eventually facilitates the lone pair activity within Pb, which further displaces the magnetic Mn motif in the system. As a result the system shows a first order structural phase transition from $P$321 to $P$3 symmetry. The presence of extra anomaly in temperature dependent XRD and dielectric data (near 120 K) is indicative of further structural disruption locally with decreasing temperature within trigonal symmetric space group of PTMPO compound. The phase transition at 310 K also affects the distribution of moments on the involved atoms locally and as a result a finite magnetoelectric coupling is observed near room temperature. Our result not only introduces a new multiferroic system with magnetoelectric coupling near room temperature but also indicate towards new mechanisms for the development of such functionalities.

$Acknowledgment$. RAS thanks CSIR, India for a fellowship. SR thanks Technical Research Center of IACS. SR also thanks Department of Science and Technology (DST) [Project No. WTI/2K15/74], UGC-DAE Consortium for Research, Mumbai, India [Project No. CRS-M-286] for support. AH and TSD acknowledge the computational support of Thematic Unit of Excellence on Computational Materials Science, funded by Nano-mission of Department of Science.

\newpage
\begin{table*}
\caption{Bond lengths Of $P$321 and $P$3 structure.}
\resizebox{14cm}{!}{
\begin{tabular}{| c | c | c |}
\hline Bond length (\AA) & Bond length of Longer site (\AA) & Bond length of Shorter site (\AA) \\
($P$321 Space group) & ($P$3 space group) & (P3 space group)\\\hline
 Mn-O = 2.01 & Mn-O = 2.11 & Mn-O = 1.91\\
 O-Pb = 2.35 & O-Pb = 2.46 & O-Pb = 2.33 \\
 O-Te = 1.93 & O-Te = 1.97 & O-Te = 1.83 \\
 \hline
\end{tabular}
}
\end{table*}
\begin{table*}
\caption{Magnetic moment of PTMPO compounds in $\mu_B$.}
\resizebox{8cm}{!}{
\begin{tabular}{| c | c | c | c | c | c |}
\hline Space group & Mn &  Pb & Te & P & O \\\hline
$P$321 & 4.632 & 0.008 & 0.034 & 0.009 & 0.021 \\
$P$3 & 4.602 & 0.012 & 0.038 & 0.013 & 0.023 \\
\hline
\end{tabular}
}
\end{table*}

\begin{figure}
\resizebox{8.6cm}{!}
{\includegraphics[89pt,359pt][515pt,718pt]{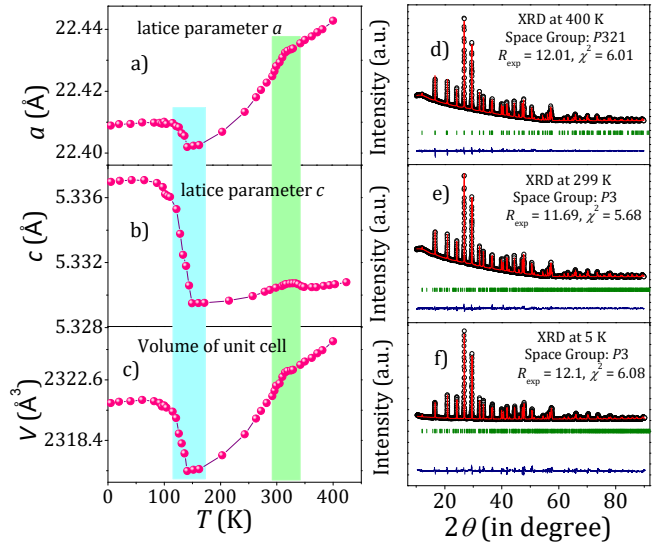}}
\caption{a), b) and c) Thermal variation of lattice parameters and volume of unit cell of PTMPO. d), e) and f) are the refined XRD pattern of 400 K, 299 k and 5 K data respectively. Open black circles represent the experimental data and continuous red line represents the calculated pattern. The blue line represents the difference between the observed and calculated pattern. Green scattered line are Bragg peak.}
\end{figure}

\begin{figure}
\resizebox{8.6cm}{!}
{\includegraphics[61pt,218pt][520pt,776pt]{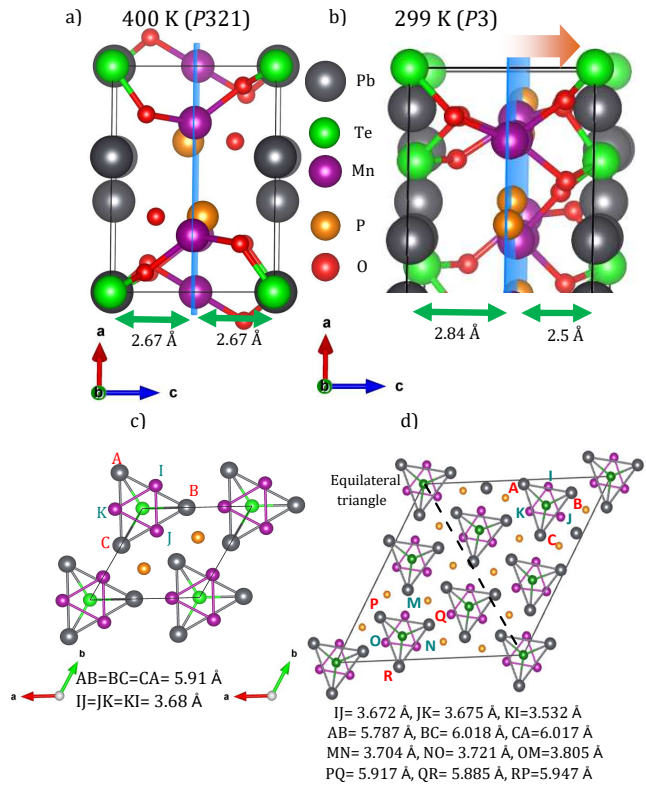}}
\caption{a) and b) are the $a-c$ plane of $P$321 and $P$3 structure respectively. c) and d)$a-b$ plane of $P$321 and $P$3 structure respectively}
\end{figure}

\begin{figure*}[h]
\resizebox{12cm}{!}
{\includegraphics[12pt,99pt][590pt,790pt]{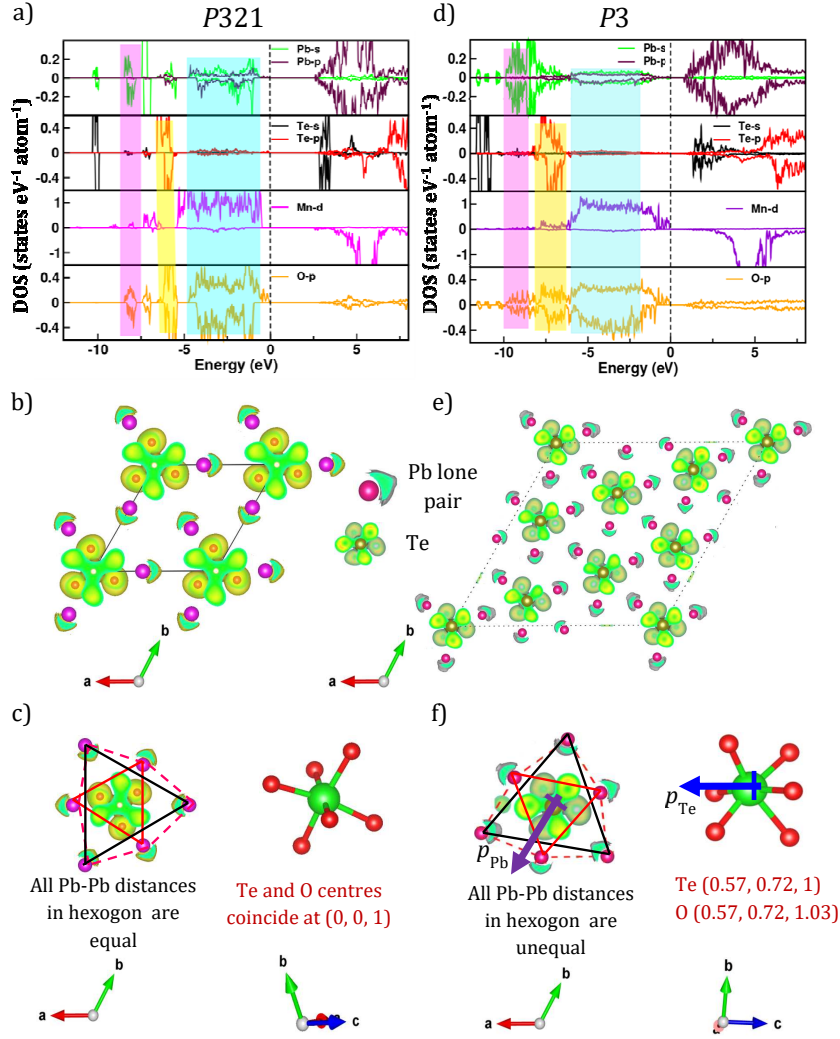}}
\caption{a), b) and c) Partial DOS, Electron localization function within a unit cell(The  isosurfaces are vizualized for a value of 0.3) and Pb$_6$ hexagon and TeO$_6$ octahedra of $P$321 structure of PTMPO respectively. d), e) and f) Partial DOS, Electron localization function within a unit cell (The  isosurfaces are vizualized for a value of 0.3) and Pb$_6$ hexagon and TeO$_6$ octahedra of $P$321 structure respectively.}
\end{figure*}

\begin{figure}
\resizebox{8.6cm}{!}
{\includegraphics[49pt,502pt][508pt,759pt]{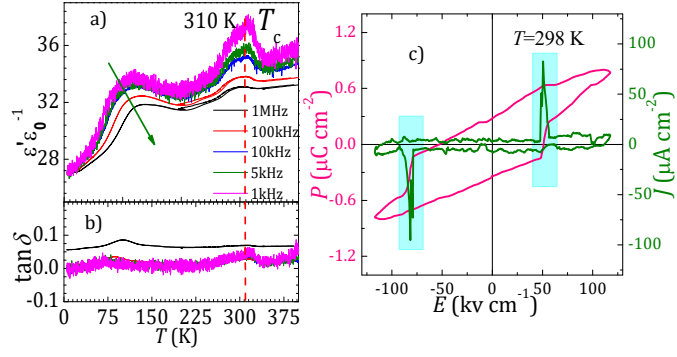}}
\caption{a) and b) Temperature dependence of real part of dielectric constant $\varepsilon'$/$\varepsilon_0$ and $\tan\delta$ loss data of PTMPO at different frequencies. c) Electric field dependent polarization (pink line) and switching current behaviour (green line)of PTMPO at room temperature.}
\end{figure}

\begin{figure}
\resizebox{8.6cm}{!}
{\includegraphics[75pt,198pt][517pt,783pt]{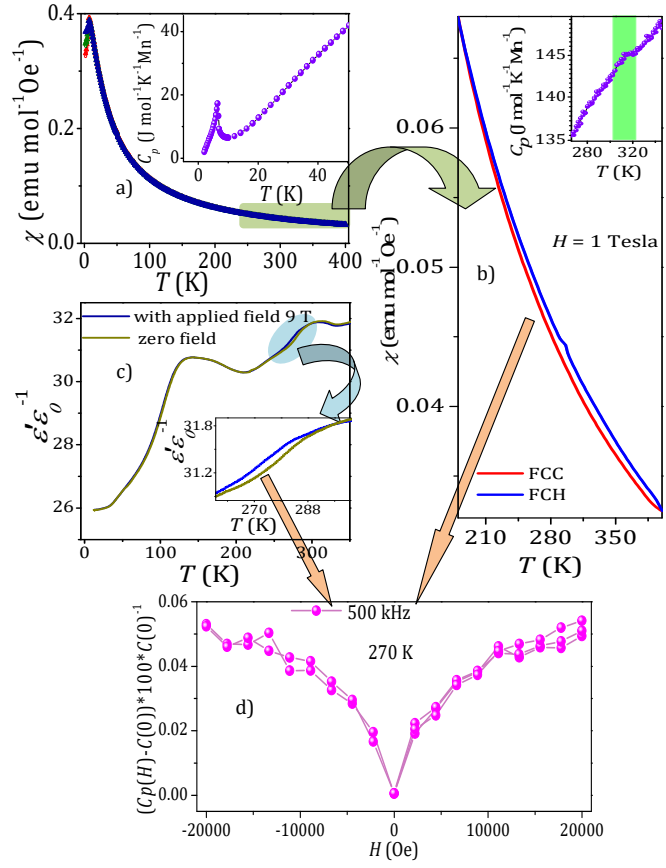}}
\caption{a) $M-T$ at 500 Oe, 5000 Oe and 10000 Oe of PTMPO. Open red, solid red triangle and solid red circle represent the ZFC, FCC and FCH of 500 Oe data. Open green, solid green triangle and solid green circle represent the ZFC, FCC and FCH of 5000 Oe data. Open blue, solid blue triangle and solid bule circle represent the ZFC, FCC and FCH of 10000 Oe data. Inset of a) shows Heat capacity data at low temperature. b) Red curve and blue curve are the FCC and FCH at 10000 Oe data. Inset of b) shows Heat capacity data at high temperature. c) Temperature dependent dielectric constant with 0 field and 9 Tesla field and inset indicates the gap between them. d) Magnetodielectric data at 270 K.}
\end{figure}

\end{document}